# Wetting film dynamics and stability

Boryan Radoev[1], Klaus W. Stöckelhuber[2], Roumen Tsekov[1] and Philippe Letocart[2]
[1]Department of Physical Chemistry, University of Sofia, 1164 Sofia, Bulgaria
[2]MPI Research Group for Colloids and Surfaces, TU Bergakademie, 09599 Freiberg, Germany

Although the wetting films are similar in many aspects to other thin liquid films, there are some differences in their behavior, too. In contrast to soap and emulsion films, whose surfaces are homogeneous, solid substrates of wetting films are heterogeneous as a rule, unless special measures for their homogenization are taken. Here we mean primarily heterogeneous distribution of surface energy leading to existence of hydrophobic domains on hydrophilic surfaces and vice versa. As is known, such hydrophobic domains could play the role of gas-phase nucleation centers and it is widely accepted nowadays that nano-bubbles can be formed there. The present paper reviews the effect of nano-bubbles adhered at solid surface on stability of wetting films. It is shown that the existence of nano-bubbles is crucial for the lifetime of wetting films. Another peculiarity typical for hydrophobic solid surface, the so-called slippage effect, is also investigated and its contribution to the dispersion equation of capillary waves on wetting films is accounted for.

Thin films, the building blocks of dispersed systems, are present mainly in three variants: soap, emulsion and wetting films. There are many equilibrium and dynamic characteristics typical for all these kinds of films, e.g. the surface DLVO forces acting therein, the type of hydrodynamics in the films (the so-called lubrication flow), the equilibrium and stability conditions, film rupture, etc. However, besides their common nature each kind of films mentioned above possesses specific features, which should be taken into account when interpreting the corresponding experimental data and their influence on the behavior of dispersed systems. The subject of this paper is to review some of the peculiarities, related to the rheology, stability and rupture of wetting films. As is shown, all phenomena observed are due to effects characteristic of solid/liquid surfaces. A section is devoted to the hydrodynamics of wetting films with special attention paid to a new experimental method for exciting and measuring surface waves on wetting films. Among the most interesting results obtained by this method is the anomalous high slippage observed on solid surfaces. Presence of nano-bubbles adhered at solid/liquid surface is suggested as a possible origin of this effect. The idea of the existence of nano-bubbles in wetting films appears again in the last section as a cause for the film rupture. The detailed analysis shows that the region above a nano-bubble resembles more a foam film, which is unstable as a rule, thus explaining why wetting films with repulsive interactions could also rupture.

# Drainage of liquid films and dimple relaxation

During the past century the hydrodynamic behavior of thin liquid films has been extensively studied in relation to the stability of disperse systems. The film drainage is usually described by the Navier-Stokes equations, applied to a liquid film of thickness much smaller than its radius. An interesting effect for thinning films is a coupling of the film drainage and the unstable hydrodynamic modes [1-3], which theoretical description requires correct calculation of the film drainage rate. The later depends substantially on the shape of the film interfaces; usually a dimple [4-6] or wimple [7, 8] form with maximum or minimum of the film thickness in the film center, respectively. Classically, the rate of thinning is described by the Reynolds equation

$$V_{Re} = \frac{2h^3 \Delta p}{3\eta R^2} \qquad (1)$$

applied first to thin liquid films by Scheludko [9]. This equation is valid for films with tangentially immobile and plane-parallel surfaces, where $h$ is the average film thickness, $R$ is the film radius, $\eta$ is the liquid viscosity, and $\Delta p = p_\sigma - \Pi$ is the driving pressure ($p_\sigma$ is the capillary pressure in the meniscus and $\Pi$ is the disjoining pressure).

Deviations from Eq. (1) due to tangential surface mobility are reported [10-14] and the theory predicts a thinning rate according to the Reynolds formula with a reduced effective viscosity $\mu$ [15-18]. In the case of wetting film the relative viscosity can be expressed as [23]

$$\mu/\eta = (Ma + Na + Ap + MaNa + ApNa/2)/(12 + 4Ma + 4Na + MaNa) \qquad (2)$$

where $Ma \equiv ahE_G / \eta(D_s a + Dh)$, $Na \equiv h\beta/\eta$ and $Ap \equiv -ah^2\Gamma(\partial_\Gamma \Pi)/\eta(aD_s + Dh)$ are the so-called Marangoni, Navier and adsorption-pressure numbers. Here $a \equiv \partial_c \Gamma$ is the adsorption length ($\Gamma$ is the adsorption and $c$ is the surfactant concentration), $E_G = -\Gamma \partial_\Gamma \sigma$ is the surface Gibbs elasticity ($\sigma$ is the surface tension on the film/air interface), $D_s$ and $D$ are the surface and bulk diffusion coefficients of the surfactant. The Marangoni number accounts for the effect of surfactants on the surface mobility, i.e. on the tangential velocity of fluid film/gas interface [15]. As is seen, $Ma$ depends on two factors, potential interactions via the Gibbs elasticity and dissipative characteristics as the viscosity and diffusion coefficients. At a pure single component liquid/gas interface (the so-called free surface) there is no adsorption ($Ma = 0$) and the tangential surface velocity is maximal. In this case Eq. (2) takes the form $\mu/\eta = Na/(12 + 4Na)$, since for a pure liquid $Ap = 0$ as well. The other limit $Ma \gg 1$ corresponds to a practically immobilized surface with zero tangential surface velocity. It is realized at sufficient amount of surfactant (high Gibbs elasticity) and Eq. (2) takes the form $\mu/\eta = (1 + Na)/(4 + Na)$, respectively.

The Navier number $Na$ accounts for the slippage on the film/solid interface. Discussions about the slippage phenomenon began [19] at the same time as the formulation of the Navier-Stokes equations. They are strongly related to the rheological models of inviscid ideal and viscid real fluids. According to Stokes, the adherence of fluids at interfaces (non-slip condition) is a straight consequence of their viscous nature. Ideal fluids slip along the interface because no adhesion arises there. Feynman called ideal fluids dry in his famous Lectures in Physics in order to emphasize their non-adhering nature. The current revival of the slip problem is related to advanced methods of experimental studies [20, 21] and to computer modeling, which makes it possible to describe the motion of a liquid along a solid from first principles. In the Navier number $\beta$ is the slip coefficient on the film/solid interface, but in the same time some authors use a slip length $b \equiv \beta/\eta$ ($Na = h/b$) to characterize the adhesion [22]. Usually for hydrophilic surfaces the non-slip boundary condition holds, which corresponds to infinite slip coefficient (zero slip length) and infinite Navier number, $Na \to \infty$. Finite slippage is experimentally detected for hydrophobic surfaces [22]. The limit $Na \to 0$ describes the boundary condition of an ideal fluid at solid surface (zero friction, divergent slip length $b \to \infty$) and from rheological point of view it is equivalent to $Ma = 0$ at the liquid/gas surface. The two cases ($Ma = 0$, $Na \gg 1$) and ($Ma \gg 1$, $Na = 0$) correspond to a film with an immobilized and a free interface, with $\mu = \eta/4$. The trivial case of two immobilized surface is equivalent to $Na \gg 1$ and $Ma \gg 1$ with $\mu = \eta$. The slippage effect is especially important for hydrodynamics of narrow regions, three-phase contact zones [22], thin films, etc., since $Na \sim h$, i.e. the thinner films possess higher slippage.

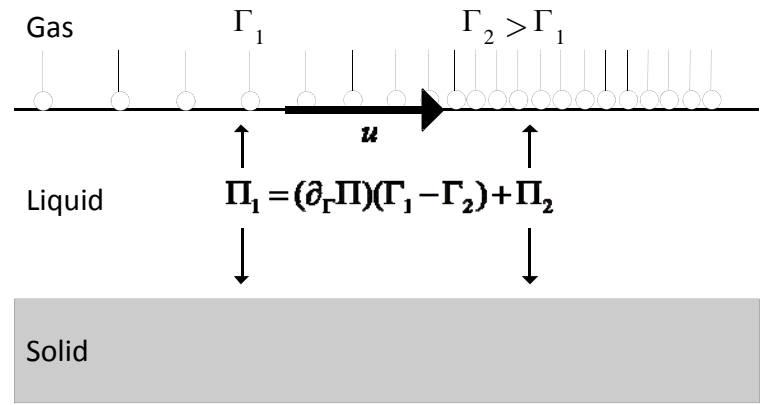

**Fig. 1** Distribution of disjoining pressure $\Pi$ due to an adsorption gradient.

Finally, the adsorption-pressure number $Ap$ takes into account the dependence of the disjoining pressure on the adsorption on the film surfaces. From this point of view it is not directly correlated with the dissipative nature of film surfaces, but rather with the driving forces in thin films. Due to the surface convection (non-zero surface velocity $u \neq 0$, see Fig. 1), a gradient of the adsorption $\nabla \Gamma$ at liquid/gas surface appears. If the disjoining pressure is a function of $\Gamma$ than it leads to a corresponding pressure gradient $\nabla \Pi = (\partial_\Gamma \Pi) \nabla \Gamma$. This effect is pronounced in the case of ionic surfactants since their adsorption gradient is equivalent to surface charge density gradient $\nabla q = F \sum z_k \nabla \Gamma_k$ ($F$ is the Faraday constant, $z_k$ is valence of adsorbed ions) and to electrostatic disjoining pressure gradient $\nabla \Pi_{EL} = (\partial_q \Pi_{EL}) \nabla q$ [70], respectively. In contrast to $Ma$ and $Na$ numbers, the $Ap$ number could be either positive or negative.

Some experiments [24] have detected rates of thinning depending much weaker on the film radius as compared to the Reynolds formula. Such a behavior cannot be explained by the surface mobility, the effect of which is not expected to change the functional dependence of the thinning rate on the film radius. Eq. (1) requires strictly symmetrical drainage between two parallel flat surfaces, which is supplied in the Reynolds case by the rigidity of the solid interfaces. In contrast, the shape of wetting films is determined via the balance of the viscous, capillary and surface forces. Experimental investigations above have shown that large films are non-homogeneous in thickness and the thinning rate of dimpled films is always larger than the prediction of the Reynolds law. Recently, a classification of the film drainage is proposed [25] by introducing a dynamic fractal dimension $\alpha$, which takes into account the effective degree of freedom for relaxation of the film interfaces. In this way a generalized drainage law is derived

$$V = \frac{2h^3 \Delta p}{3\mu R^2} (\frac{R^2 \Delta p}{16 \sigma h})^{\frac{2-\alpha}{2+\alpha}} \tag{3}$$

Eq. (3) discriminates between different types of drainage by the corresponding value of $\alpha$. For instance, if the film possesses two solid interfaces the latter induce strong correlation in both two directions and $\alpha = 2$. Hence, according to Eq. (3) the film drains with the Reynolds velocity. If the drainage is strictly axisymmetric but not confined between solid interfaces, the radial direction is free for relaxation and a dimple occurs. In this case $\alpha = 1$ and Eq. (3) reduces to

$$V = \frac{1}{3\mu} \sqrt[3]{\frac{h^8 \Delta p^4}{2\sigma R^4}} \tag{4}$$

One can imagine a completely random film broken down to many uncorrelated sub-domains. In this case $\alpha = 0$ and Eq. (3) acquires the form

$$V = \frac{h^2 \Delta p^2}{24 \mu \sigma} \tag{5}$$

Note that the rate of drainage of such stochastically corrugated films does not depend on the film radius since any spatial correlation of the dynamics is missing. Finally, one can imagine that a film is broken to sub-domains, which are not completely uncorrelated. In this case $\alpha = 1/2$, if the pressure fluctuations in the film obey the thermodynamic law [26], i.e. their amplitude is inversely proportional to the square root of the number of sub-domains in the film. Hence, Eq. (3) changes to the following rate of drainage

$$V = \frac{1}{6\mu} \sqrt[5]{\frac{h^{12}\Delta p^8}{4\sigma^3 R^4}} \tag{6}$$

This equation predicts that the rate of drainage is inversely proportional to the film radius on the power 4/5. Such a behavior was experimentally observed [24, 26]. The non-linear dependence of the thinning rate $V$ on the driving pressure $\Delta p$ from Eq. (6) is also experimentally proven [25]. In general, the dynamic fractal dimension $\alpha$ depends on the film radius [71].

In fact, the thinning of wetting films is a more complicate process passing through several stages. Since initially the drainage at the film border is much faster than that in the film center [4-6, 27, 28], the thickness at the barrier rim reaches quickly a value close to the equilibrium thickness $h_e$. This dimpled structure is non-equilibrium one and a next stage in the film thinning is the dimple relaxation. Under the action of capillary and disjoining forces the dimple shrinks till the film reaches flat shape at equilibrium. Sometimes, however, under non-equilibrium conditions the dimples can grow driven by external flows [29]. In the frames of the lubrication approximation the evolution of the local thickness $H$ of an axisymmetric wetting film is governed by the following equation [23, 30]

$$\partial_t H = -\frac{1}{r}\partial_r \{\frac{H^3}{12\mu(H)} r \partial_r [\frac{\sigma(H)}{r} \partial_r (r \frac{\partial_r H}{\sqrt{1+(\partial_r H)^2}}) + \Pi(H)]\} \equiv \hat{L}\{H\} \tag{7}$$

where $t$ is time, $r$ is radial coordinate and $\hat{L}$ is the evolution operator. Due to complexity of this equation it could be solved only numerically [29]. Hereafter we present an approximate procedure for description of the dimple relaxation which elucidates the physical aspects and is more transparent. A rigorous treatment of the dimple relaxation via Eq. (7) requires relevant boundary and initial conditions. The exact boundary conditions can be written only far away in the meniscus, where the lubrication theory, respectively Eq. (7), is no longer applicable. Moreover, Eq. (7) is valid for the latest stage of the dimple relaxation and the relevant initial profile can only be experimentally specified. Hence, it is impossible to close the mathematical problem and the further treatment requires an empirical modeling. For this reason, an approximate method is developed [30, 31] for calculating the evolution under the assumption that the film shape is known.

The film profile can be generally represented as a $r^2$-power expansion. Since $r \leq R$, one is able to approximate the dimple in the film region by a finite series, e.g. a biquadratic form. It can be additionally specified by application of the two known conditions at the barrier rim: equilibrium thickness $H(R) = h_e$ and minimum of the film profile $(\partial_r H)_R = 0$. Thus, the biquadratic polynomial form acquires the formula

$$H = h_e + [h_0(t) - h_e][1 - (r/R)^2]^2 \tag{8}$$

The thickness $h_0$ in the film center is the only unknown function of time in Eq. (8). Suppose, at time $t$ the film profile is given by Eq. (8). The profile at time $t + \tau$ can be calculated from Eq. (7), which in the limit of small $\tau$ can be rewritten as

$$\tilde{H}(r, t + \tau) = H(r, t) + \tau \hat{L} H(r, t) \tag{9}$$

Thus, the consequent profile $\tilde{H}$ can be generated, which satisfies Eq. (7) but does not obey the necessary boundary conditions. To apply the boundary conditions we postulate that the real profile $H(r, t + \tau)$ is the best fit of the profile $\tilde{H}(r, t + \tau)$. Hence, to calculate the film profile a minimization of the square of the deviation of the two functions all over the film is required. This criterion combined with Eqs. (8) and (9) leads to the following recurrent relation

$$h_0(t+\tau) = h_0(t) + 10\tau \int_0^1 (1-x^2)^2 \hat{L}[h_0(t)(1-x^2)^2 + h_e x^2(2-x^2)] x \, dx \tag{10}$$

where $x = r/R$. It is only a matter of integration to calculate the evolution of the thickness in the film center, which introduced in Eq. (8) will provide the whole film profile evolution.

We have applied the method described above to wetting films from 1 mM aqueous solution of KCl on a glass surface [32]. In order to check the theory we have compared it with experimental results of dimple relaxation and a very good agreement is observed for the thinning in the film center, which is monitored by the classical light interference method. Assuming that the OH⁻ ions adsorb of the water/air interface [33], we estimated their adsorption length $a = 0.14$ mm. This is a reasonable value since the concentration of OH⁻ ions is very low and it is well-known that the slope of the adsorption isotherm is large when the concentration approaches zero. The value of $a$ corresponds to adsorption $\Gamma = 14$ nmol/m² and surface charge density $q = -1.35$ mC/m² on the water/air surface. The distribution of the OH⁻ ions on the wa-ter/air surface was also simulated. Initially, the adsorption in the film center is about 30% less than the equilibrium value due to film drainage. With advancing time the thinning rate of the film is going down and the adsorption increases to reach its equilibrium value at the end of the process. Thus for film thickness larger than 80 nm $Ap$ is small and the reduced viscosity $\mu$ practically equals to $\eta$. If the film thickness decreases below 80 nm the adsorption-pressure number becomes negative, which decreases $\mu$ and close to the equilibrium thickness it is twice smaller than $\eta$. It is important to note here that the Marangoni number $Ma$ of the OH$^-$ surfactant in water is about 1000. Hence, the water/air surface is tangentially immobile even in the case of pure water. This is important novelty since in hydrodynamics the water/air surface is usually considered as a free one with zero stress.

Our next aim was to investigate further the effect of ionic surfactants on the film dynamics [34]. For this reason, wetting films were formed on a glass surface from 1 mM aqueous solution of sodium dodecyl sulfate (SDS) [30], which is an anionic surfactant. SDS contributes to the dimple evolution either through the Marangoni effect or through the disjoining pressure; the adsorption of DS$^-$ ions generates electrostatic disjoining pressure much larger than the van der Waals component at the considered concentration [35]. The evolution of the thickness at the film center and the thickness at $r = 0.625R$ was monitored by the classical light interference method again. By comparing the theory with experimental data the surface diffusion coefficient of DS$^-$ ions is determined as a single fitted parameter. Thus, important information about the interfacial properties of DS$^-$ adsorption layer is obtained. It is shown that the interfacial electrostatics decreases dramatically the Marangoni number by suppressing the surface elasticity. This effect enhanced by the effect of the adsorption dependence of the disjoining pressure leads to important increase of the thinning rate at small film thickness. For film thickness larger than 100 nm $Ap$ is small and negative and $Ma$ is large and positive. However, if the film thickness drops below 100 nm, these numbers decrease strongly. Therefore, the coupling between the disjoining pressure and adsorption is important and leads to substantial increase of the thinning rate at the end stage of the film drainage. It could be a plausible explanation of the discrepancy in the mobility determined by drainage experiments and by dynamics of artificial waves [35] on an equilibrium film from ionic surfactant solution.

## Acoustically exited surface waves

Capillary surface waves on liquids have been investigated for a very long time beginning with the first prediction by Smoluchowski in 1908 that a liquid surface must scatter light [36], mainly near the critical temperature. The first quantitative theory was developed by Mandelstam in 1913, who obtained the root mean square amplitude $A_\zeta$ of surface waves on deep water on the basis of the equipartition theorem [37]

$$\sigma A_\zeta^2 \sim k_B T \tag{11}$$

Here $k_B$ is the Boltzmann constant, $T$ is the absolute temperature and $\sigma$ is the surface tension. Systematic study of capillary waves on thin film surfaces began in the late 60s of the previous century and since then a great number of relevant publications exists [38-43]. Artificial wave propagation in foams was studied specifically by Sun et al. [44], bending mode in free films by Bergmann [45] and wetting films by Schulze et al. [23, 35].

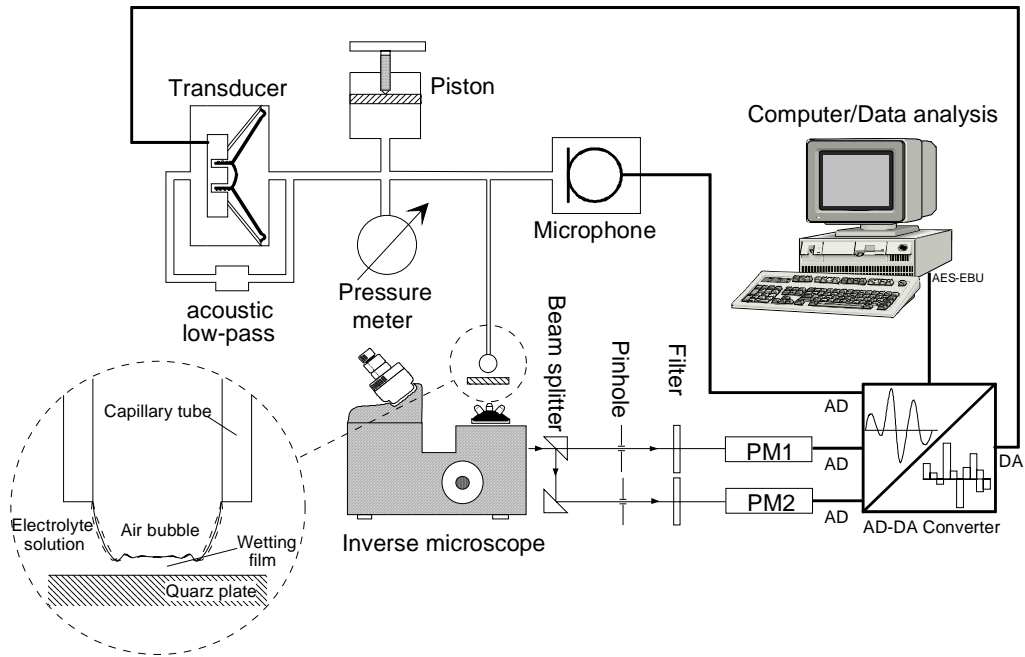

**Fig. 2** Overview of the Derjaguin-Scheludko force balance (DSFB) apparatus for studying of acoustically exited waves on wetting films; frequency range of oscillation is 10 to 200 Hz. Zoomed inset: oscillating film with the adjacent meniscus (bubble surface).

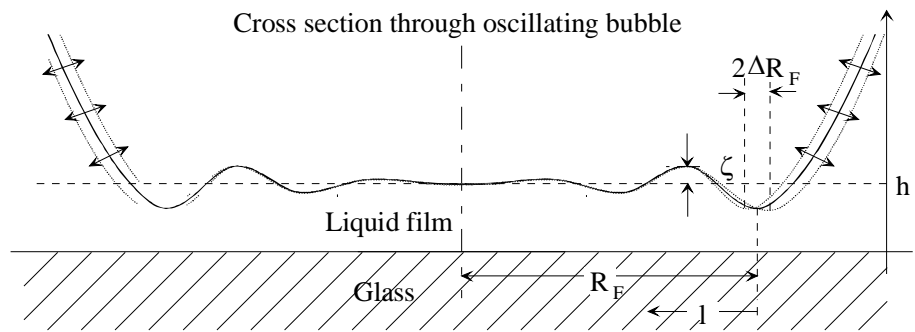

**Fig. 3** Cross-section of oscillating bubble (meniscus) and wetting film. Note that the meniscus and the film are mechanically coupled.

The core method presented here, the so-called Derjaguin-Scheludko force balance (DSFB) [35] for exciting and measuring surface waves on wetting films, is shown in Figs. 2 and 3. The equilibrium thickness of the film is adjusted by a piston and monitored by a baratron gauge. The dynamic pressure variations are generated by a loudspeaker. By this way, a wave with a circular wave front is propagating from the film edge towards film center (Fig. 3).

As is known, a dynamic fingerprint of a wave process is its dispersion relation $\omega(k)$ with $\omega = 2\pi\nu$ ($\nu$ is the wave frequency) and $k = 2\pi/\lambda$ ($\lambda$ is the wave length). For long waves in wetting films ($\lambda \gg h$) the following dispersion relation is derived [23]

$$\omega = ih^3 k^2 (\sigma k^2 - \partial_h \Pi)/3\mu \tag{12}$$

Actually, Eq. (12) is an analogue to Eq. (1) since $\omega$ is proportional to the velocity $\partial_t \zeta = i\omega\zeta$ and $(\sigma k^2 - \partial_h \Pi)$ plays the role of driving pressure corresponding to $\Delta p$. It is essential that $\mu$ in Eq. (12) is the effective viscosity defined in Eq. (2), which provides the principle opportunity to get information about the thermodynamics and rheology of wetting films. As is seen below, this possibility can be realized by measuring $\lambda$ vs. $\nu$ via the DSFB method. The results for this model system are summarized in Fig. 4:

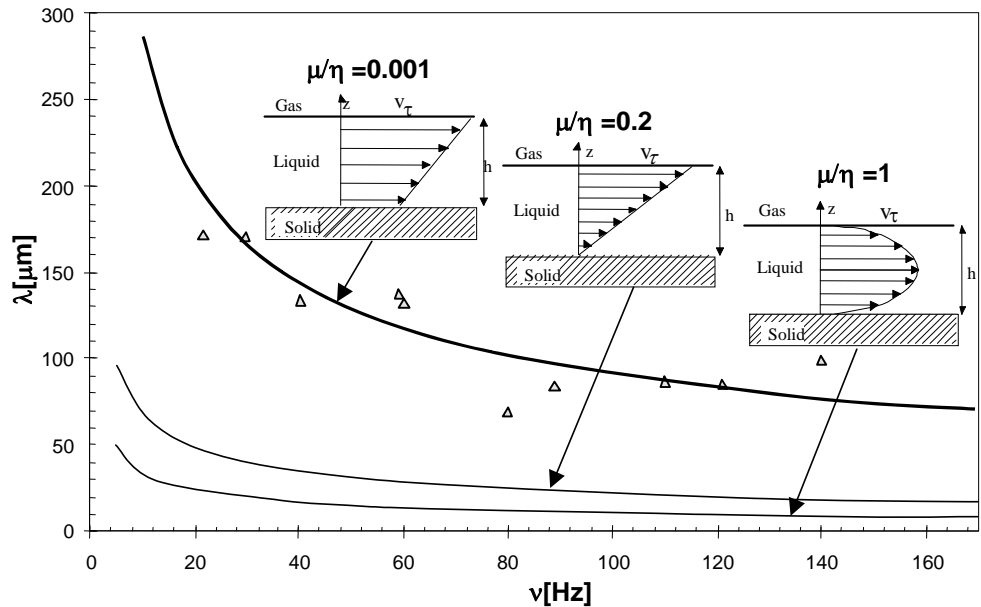

**Fig. 4** Dispersion equation $\lambda$ vs. $\nu$; $\Delta$ - experimental points for a model system glass/liquid/gas with $h_e$ = 80 nm, $\partial_h \Pi$ = 19 N/mm$^3$, $\psi_1 = -80$ mV, $\psi_2 = -100$ mV. The solid lines are theoretical dependence for three different $\mu$ values from Eq. (12). The insets illustrate the slip conditions to the corresponding $\mu$ values.

- Due to small concentration of surfactant, both Marangoni and adsorption-disjoining pressure effects are negligible;
- The relation between the frequency and the wave length of this model system corresponds qualitatively with the theory from Eq. (12) at $\sigma k^2 << |\partial_h \Pi|$;
- The traditional assumption $Na \to \infty$ on solid/liquid surface as a fully blocked surface cannot be maintained. On the contrary, a value of $Na = 1/5$ is required in order to fit the experimental results.

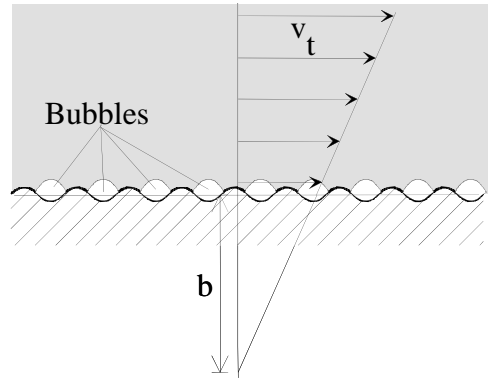

**Fig. 5** A model of enhanced solid/liquid mobility due to adhered nano-bubbles; $b$ is the so-called slip length; $v_t$ is the tangential velocity.

Experiments were carried out for wetting films of different aqueous solutions of potassium chloride (0.1 to 1 mM) on quartz, with small amount (0.1 mM) of sodium dodecyl sulfate (SDS) as an air/electrolyte surface potential stabilizer. The results on Fig. 4 show that this method reveals interesting surface rheology effects [23, 35]. The experiment shows that there exists a quantitative discrepancy with the traditional non-slip condition at solid surfaces, which can only be explained by a more generalized rheological model. A possible origin of the established slippage could be nano-bubbles adhered at the solid interface (see Fig. 5). As discussed above, there are many evidences for the existence of nano-bubbles at not-perfectly wettable water/quartz interface. On Fig. 5 it is shown schematically how the role of the finite bubble liquid/gas surface mobility leads to a pseudo-slip effect. The consequence is the obtained anomalously low $Na$ number.

**Stability of wetting films**

Stability and rupture behavior of aqueous wetting films on solid surfaces is an important step in lots of coagulation processes of colloidal systems. Particularly, the interaction of an air bubble with a solid particle in the industrial flotation process [46], which is widely used in min-eral processing, in paper recycling or waste water treatment, the forming, thinning and rupture of thin intervening water films play a crucial role. Whether the film rupture takes place in the short contact time of the particle and the bubble (millisecond scale) or not is of decisive consequence for flotation success and yield.

In general, the positive sign of the disjoining pressure $\Pi > 0$ is the necessary condition for equilibrium liquid films (see Figs. 6 and 7), whereas their stability depends on the sign of the disjoining pressure derivative $\partial_h \Pi < 0$ [9]. Usually, the disjoining pressures increase in absolute value by decreasing thickness and then both the equilibrium and stability conditions can be determined only by its sign: $\Pi > 0$ (equivalent to $\partial_h \Pi < 0$) guarantees equilibrium and stable films and vice versa, $\Pi < 0$ (equivalent to $\partial_h \Pi > 0$) corresponds to unstable films (see Figs. 11-13). Experiments with foam and emulsion films mainly [16] have shown excellent agreement with this theory, while wetting films have revealed some peculiarities. Further it will be shown that part of these peculiarities is due to substantial heterogeneity typical for solid surfaces.

The most investigated systems in this field are aqueous wetting films on silica surfaces; i.e. quartz, glass or the oxide layer on silicon wafers. In this asymmetric system all DLVO-forces are repulsive, $\Pi_{VW} = -A/6\pi h^3 > 0$ since $A = -1 \times 10^{-20}$ J [47] and $\Pi_{EL} > 0$ since the potential of the clean water/air interface is negative $\psi_1 = -35$ mV [48] and the silica/water interface at neutral pH-value is negatively charged $\psi_2 = -30$ mV, too. Thus, stable wetting films can be formed on meticulously cleaned silica surfaces. Figure 6 presents a typical picture of a stable equilibrium film formed in DSFB of 1 mM KCl solution on hydrophilic silica. The uniform gray inside the wetting film shows a homogeneous thickness of the film. The Newton fringes at the film edge are due to the meniscus profile at the bubble film contact.

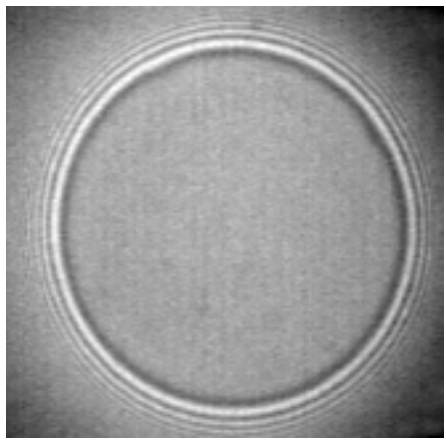

**Fig. 6** Flat stable equilibrium film of 1 mM KCl solution on hydrophilic silica.

On Fig. 7 DLVO-forces versus distance curves are represented for three different electrolyte solutions. As is seen, the corresponding equilibrium thicknesses $h_e$ are simply the abscise coordinates of the intersection point between the total disjoining pressure $\Pi$ and the capillary pressure $p_\sigma$ in the meniscus, equivalent to the pressure balance $\Delta p = p_\sigma - \Pi(h_e) = 0$ in Eq. (1). It should be pointed out here again that the case of both repulsive van der Waals and electrostatic double layer forces is limited to some asymmetric films and can therefore be only found in ternary systems like gas/water/solid or oil/water/solid [47]. In colloidal systems where interact two identical media trough a second phase, the van der Waals forces are always attractive, so only metastable films can be found. The above mentioned technological flotation process does not succeed, if stable water films are forming between solid particles and gas bubbles.

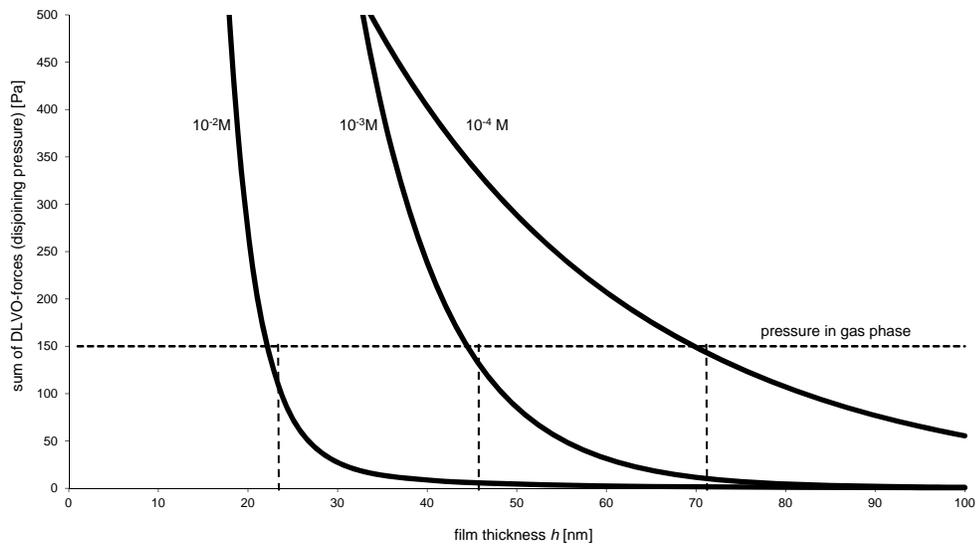

**Fig. 7** Force/distance curves for wetting films on silica with ionic strength 0.1÷10 mM KCl and $A = -1 \times 10^{-20}$ J, $\psi_1 = -35$ mV, $\psi_2 = -30$ mV. The line gives a typical DSFB experimental bubble pressure. The abscise coordinates of the intersection points between the disjoining pressure $\Pi$ and the capillary pressure $p_\sigma$ in the meniscus correspond to equilibrium thicknesses $h_e$.

### Rupture of wetting films (unstable wave mechanism)

If the resultant DLVO-force in the film is attractive ($\Pi < 0$), the most important consequence is that the film becomes unstable, i.e. its drainage ends inevitably with a rupture. We have studied this rupture process in systems of repulsive van der Waals force ($\Pi_{VW} > 0$) and attractive electrostatic interactions ($\Pi_{EL} < 0$) so that the total interaction forces remains attractive ($\Pi_{VW} + \Pi_{EL} < 0$). The occurrence of attractive electrostatic forces can be experimentally realized by overcharging the silica surface by $Al^{3+}$ ions [49, 50] or by applying TiO$_2$ on glass-surfaces in a pH-range more acid than the isoelectric point of this system (own measurements, still unpublished). Figure 8 shows $\Pi(h)$ isotherms of a silica surface overcharged by $Al^{3+}$ ions. As seen, the electrostatic interaction for this particular system predominates in the entire interval above rupture thicknesses ($h_{cr} = 35$ nm, see Fig. 9) and the resultant disjoining pressure is practically equal to $\Pi_{EL}$ [53]

$$\Pi_{EL} = 2\varepsilon_0\varepsilon\kappa^2(4k_BT/ze)\tanh(ze\psi_1/4k_BT)\tanh(ze\psi_2/4k_BT)\exp(-\kappa h) \qquad (13)$$

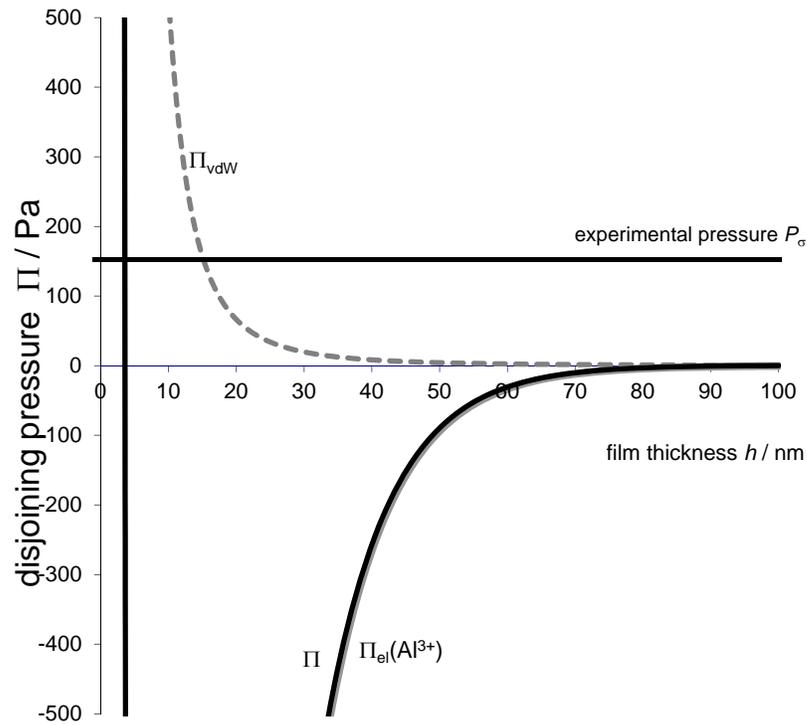

**Fig. 8** Force distance curves for films containing 0.1 mM $AlCl_3$ and 1 mM KCl. The solid surface is positively charged due to the $Al^{3+}$ ions ($\psi_2 = +35$ mV), the air/liquid interface remains negatively charged ($\psi_1 = -35$ mV). $\Pi = \Pi_{VW} + \Pi_{EL}$ is the total disjoining pressure.

As was already mentioned, fluid surfaces are corrugated due to thermal fluctuations but in contrast to deep water, where the waves are stable with finite amplitudes Eq. (11), surface waves in thin films could be unstable with increasing amplitudes. In the case of attractive DLVO-forces, a principle possibility arises that part of the wave spectrum may become unstable. According to the theory [9, 40], a wave named critical wave with a length $\lambda_{cr}$ splits the Fourier spectrum into stable and unstable parts (Fig. 9). The critical wavelength is proportional to the square root of the film surface tension divided by the first derivative of the disjoining pressure with respect to the film thickness [9]

$$\lambda_{cr} \sim \sqrt{\sigma / \partial_h \Pi} \qquad (14)$$

Relation (14) follows directly from Eq. (12) by substituting $\sigma k^2 = \partial_h \Pi$ therein. Fig. 9 represents a typical trend of the function $\lambda_{cr}(h)$ calculated from Eq. (14) with $\Pi$ from Eq. (13). As seen, the short-waves branch $\lambda < \lambda_{cr}$ is stable, while the long-waves branch $\lambda > \lambda_{cr}$ is unstable.

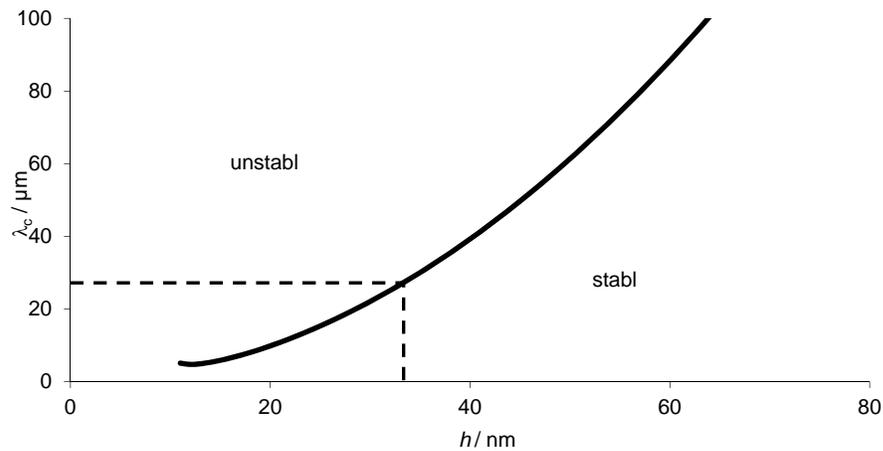

**Fig. 9** Stable and unstable capillary wave spectrum regions; $\lambda_{cr}(h)$ curve calculated by Eqs. (13) and (14) for the experimental conditions of Fig. 8. The particular point ($h_{cr} = 35$ nm, $\lambda_{cr} = 25$ µm) is in a good agreement with experimental observations from see Fig. 10.

It is worth noting that the unstable modes rupture mechanism considered here is equivalent to a spontaneous non-equilibrium process, i.e. unstable films will rupture at any stationary thickness $h = const$ for which a critical mode $\lambda_{cr}(h)$ exists. In the next section we will see that for heterogeneous wetting films there is also a possibility for another mechanism, where rupture occurs only at a definite thickness via a non-spontaneous mechanism, Eq. (18). When the amplitudes of the unstable waves are increasing, they will reach the opposite film surface at a given moment. For soap films this contact point usually leads to rupture [40], while for wetting films this is followed by a hole-formation, known in the literature also as dewetting process [49-51]. If one assumes that every touch leads with large probability to the formation of a hole, the distance between them should be scaled by the wave length. The experimental data from Fig. 10 are in a good agreement with this prediction. The evaluated $\lambda_{cr}$ value from Fig. 9 for the experimental measured rupture thickness $h_{cr} = 35$ nm leads to a critical wave length of approximately 25 µm, which is of the same order as the measured distance of the holes in Fig. 10.

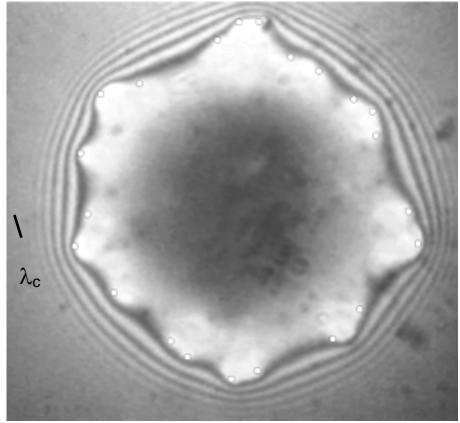

**Fig. 10** Ruptured film on silica, positively charged by $10^{-4}$ mol/l $Al^{3+}$-ions; $10^{-2}$ mol/l KCl electrolyte. The holes in the film marked as white dots for better visibility are nearly equidistant

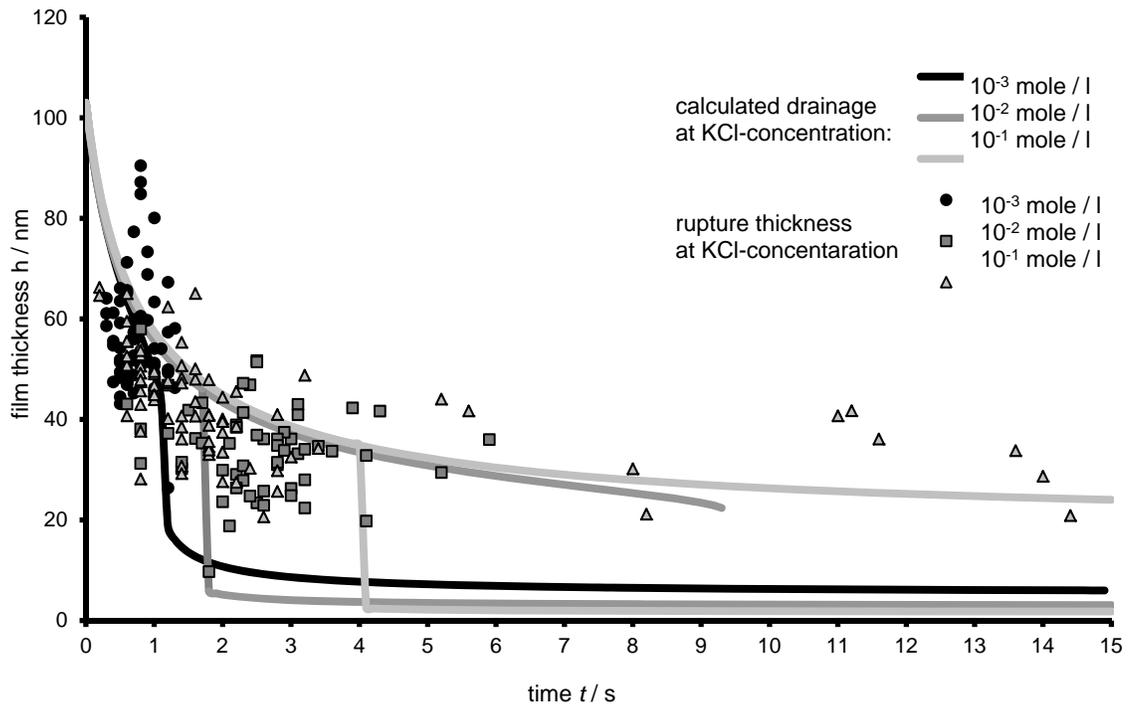

**Fig. 11** Drainage and rupture of a wetting film of 0.1 mM $AlCl_3$ aqueous solution on glass at different KCl concentrations. The curves are calculated from the Reynolds law (1) and the dots represent the rupture thickness/lifetime.

Another evidence for the existence of the capillary wave rupture mechanism can be seen in the results of Fig. 11. The curves represent film drainage calculated by Eq. (1) at three different KCl concentrations, i.e. at three different electrostatic disjoining pressures (see Fig. 7). As is seen, the experimentally measured rupture thicknesses (the different types of dots) lay in the range of the corresponding Reynolds law curves [52, 53]. The high scattering of the rupture data is typical for metastable systems.

## Film rupture caused by nano-bubbles (nucleation mechanism)

Let us consider aqueous wetting film on pure silica surface, i.e. no recharged by $Al^{3+}$ ions. As already discussed, such films are stable with equilibrium thicknesses depending on the KCl concentration and the experimental pressure $p_\sigma$ (see Fig. 7). But if the silica surface is made hydrophobic, e.g. via methylation, wetting films become unstable and rupture, although the DLVO-forces remain repulsive. It should be recalled that the hydrophobization does not change the repulsive nature of DLVO-forces, as far as neither the Hamaker constant nor the surface potential at neutral pH-values of the silica surface are changed significantly [54, 55]. The rupture thickness in this system can reach very high values, up to hundreds of nanometers, whereas the lifetimes are very short, from the order of parts of a second. To explain this problem, some groups introduced a long range hydrophobic force of unknown physical origin [56, 57]. Nowadays it is widely accepted that the rupture of these films is caused by small gas bubbles in sub micrometer range adhered at the hydrophobic solid surface [58]. A number of papers are published that give clear evidence for the existence of such nano-bubbles by means of IR-spectroscopy [59], force measurements [60-62] and also by image-giving methods like tapping-mode atomic force microscopy [63-65]. As already mentioned, the rupture mechanism in this case is non-spontaneous, in contrast to the spontaneous unstable wave mechanism. However, due to the way of bubble formation, it is known in the literature as nucleation rupture mechanism [50-53]. Figure 12 shows rupture thicknesses and lifetimes, respectively, of wetting films at different hydrophobized silica surfaces. The degree of hydrophobization is characterized by the corresponding advancing angle; 20, 58 and 90 $^0$ in this particular case. The following conclusions can be drawn from the experimental data presented in Fig. 12:
- The larger the degree of hydrophobization (the larger the advancing contact angle), the higher the rupture thickness (the shorter the lifetime);
- The ruptures take place along the theoretical drainage curve from Eq. (1), where only repulsive DLVO- forces are taken into account.

The last observation strongly supports the assumption mentioned above that the hydrophobization does not change the repulsive character of the surface forces in the wetting film on silica surface and that no additional attractive forces (like long range hydrophobic forces) are acting in this case.

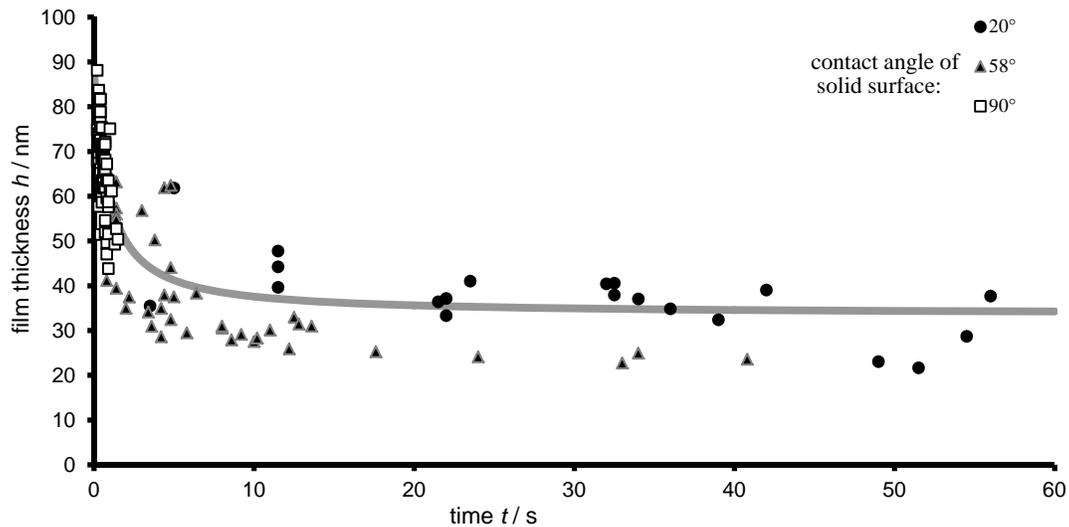

**Fig. 12** Drainage and rupture of wetting films on methylated silica with different contact angles. Line is calculated curve according to Reynolds law (1 mM KCl); dots are rupture thickness.

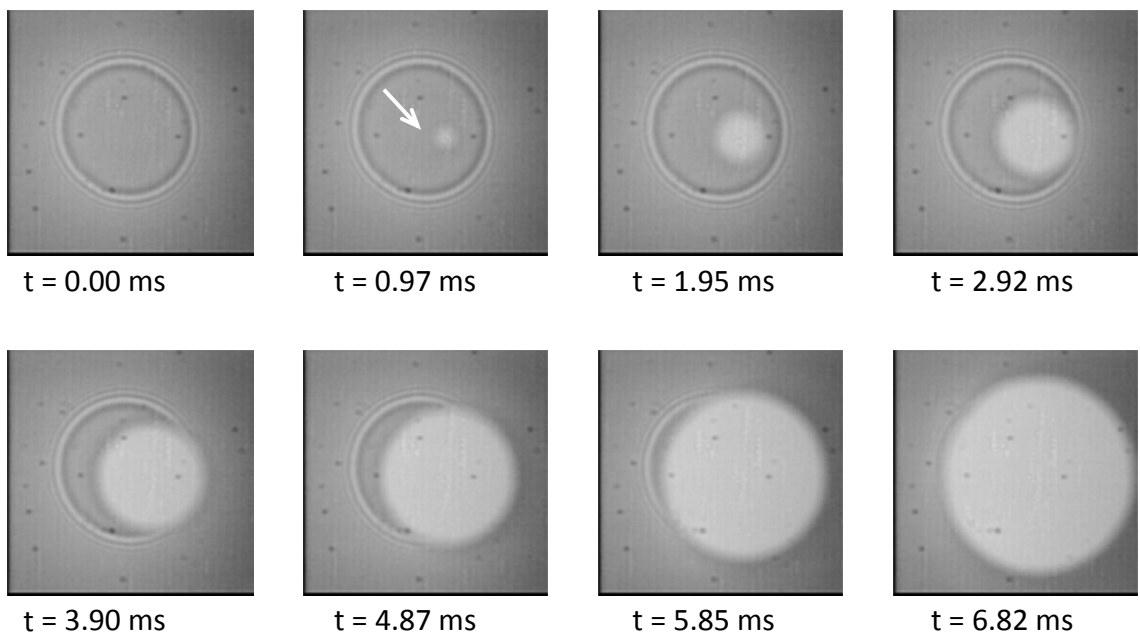

**Fig. 13** Microinterferometric high speed video frames of the rupture of an aqueous wetting film on methylated silica. The arrow on the second frame shows the point, where nucleation starts.

Also the microscopic observed rupture (Fig. 13) gives an indication of a nucleation process: the rupture starts from one single hole somewhere in the film, in all probability at the place where the biggest nano-bubble is located, and leads to a complete dewetting of the whole film area (radius ca. 100 µm) in a few milliseconds. Note the drastic difference with the film evolution in the case of hydrophilic silica surfaces where ruptures occur at several points but the contact spots do not expand (Fig. 10). The detailed analysis [52, 53, 66, 67] shows that gas bubbles formation actually creates non-homogeneities in the wetting film. In the region above the bubble the surface forces are much closer to surface forces in foam (air/water/air) films and differ from those in wetting films (Fig. 14). Wetting films (silica/water/air) have both van der Waals and electrostatic forces repulsive (Fig. 12), while in foam film the van der Waals interactions are attractive. In Fig. 14 are given schematically the equilibrium profile and the pressure balance of a wetting film caused by a bubble adhered at the solid surface.

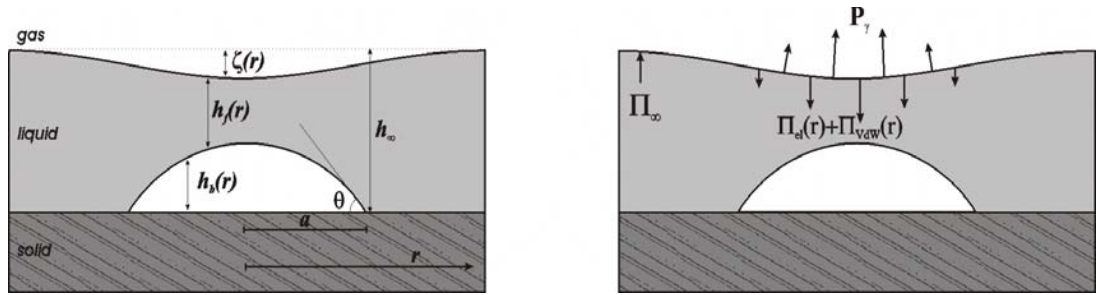

**Fig. 14** Equilibrium profile and pressure acting on a film surface in presence of a bubble; $h_\infty$ and $\Pi_\infty > 0$ are the wetting film thickness and repulsive disjoining pressure far from the bubble.

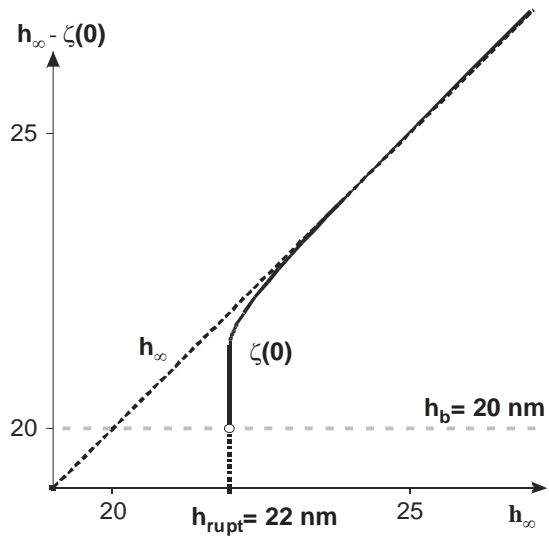

**Fig. 15** Equilibrium deformation of a film surface above a bubble apex (see Fig. 14). The rupture occurs when $\zeta(0)$ reaches the bubble apex. Rupture thickness value $h_{cr} = 22$ nm computed with Eq. (15) is close to the bubble height of 20 nm.

Quantitative results of the film surface deformation above the bubble apex $\zeta(0)$ are demonstrated in Fig. 15. The equilibrium form $\zeta(r)$ is solution of the pressure balance [67]

$$\Pi(r,\zeta) + p_\sigma(\zeta) = \Pi(h_\infty) \tag{15}$$

where $\Pi(r,\zeta)$ and $\Pi_\infty = \Pi(h_\infty)$ are the resultant disjoining pressures inside and outside the bubble region and $p_\sigma(\zeta)$ is the local capillary pressure. In contrast to the disjoining pressure $\Pi_\infty$, which depends solely on the wetting film thickness, $\Pi$ is a function of the radial coordinate. Actually, in this case we have a multilayer (gas/liquid/gas/solid) with variable thicknesses of liquid phase $h_f(r)$ and of the bubble gas phases $h_b(r)$ (Fig. 14), which reflects on the radial dependence of $\Pi$. For example, the corresponding van der Waals disjoining pressure becomes

$$\Pi_{VW}(r,\zeta) = A_{SL}/6\pi[h_\infty - \zeta(r)]^3 - A_{LL}/6\pi h_f^3(r) \tag{16}$$

with $A_{SL} = 4.7 \times 10^{-20}$ J $> A_{LL} = 3.7 \times 10^{-20}$ J being the Hamaker constant for solid/liquid media [68]. Obviously, at sufficiently thin liquid layer ($h_f < h_\infty - \zeta$) above the bubble, the second term in Eq. (16) prevails and $\Pi_{VW} < 0$, i.e. we expect local attractive van der Waals interactions. In contrast, far away from the bubble in wetting film, the van der Waals interaction is repulsive

$$\Pi_{VW}(h_\infty) = (A_{SL} - A_{LL})/6\pi h_\infty^3 > 0 \tag{17}$$

Fig. 15 represents an equilibrium deformation of a film above a bubble apex. As is seen, the surface deformations are negligible ($\zeta < 3$ nm) except for heights shortly before the moment of contact with the bubble, i.e. just before the film rupture. According to the mechanism discussed before, films rupture when their two surfaces touch each other at least in a single point. In the system considered here this condition reads:

$$\zeta(0, h_{cr}) = h_{cr} - h_b(0) \tag{18}$$

Note that according the above definition $h_{cr}$ is the wetting film thickness (far away from the bubble), i.e. $h_{cr} = h_\infty$. For the particular case in Fig. 15 the theoretically estimated rupture thickness is close to the bubble height, i.e. $h_{cr} \approx h_b(0)$ [67]. Condition (18) defines an equilibrium (non-spontaneous) rupture mechanism, but as already mentioned, the complete solution of a rupture problem needs an additional stability analysis. Generally, stability depends on the evolution of perturbations of the equilibrium characteristics, i.e. whether the perturbation amplitudes grow or diminish with time [69]. For thin liquid films the thickness perturbation evolution $\delta h(r,t)$ is significant, i.e. at $\partial_t |\delta h| < 0$ the film is stable, while at $\partial_t |\delta h| > 0$ it is unstable. Obviously, the behavior of $\delta h(r,t)$ is determined by the surface forces, i.e. by the pressure balance perturbation $\delta(p_\sigma - \Pi)$, see Eq. (15). The relation between $\partial_t \delta h$ and $\delta(p_\sigma - \Pi)$ obeys Eq. (7). The evolution equation defining $\delta h(r,t)$ follows directly in explicit form by substituting in Eq. (7) of $H = h_e + \delta h$. After expanding the terms of the sum in a series by taking into account the equilibrium condition (15), $p_\sigma + \Pi \approx (p_\sigma + \Pi)_e + p_\sigma(\delta h) + (\partial_h \Pi)\delta h$, one obtains the following expression for the evolution equation of $\delta h$ [16]

$$\partial_t \delta h = -\nabla \cdot \{h_f^3(r)\nabla[\sigma\nabla^2\delta h + (\partial_h \Pi)_e \delta h]\}/12\mu \qquad (19)$$

For the sake of briefness the standard 2D nabla operator $\nabla$ is used in Eq. (19). Note that in the frame of the linear perturbation approach the capillary pressure $p_\sigma = -\sigma\nabla^2\delta h$ is also linearized. For homogeneous films, $h_f = const$, Eq. (19) leads after applying the standard Fourier transformation to the dispersion condition (12). The main problem with heterogeneous films is that the Fourier technique is not applicable anymore to Eq. (19) because of uneven thicknesses $h_f(r)$ and variable disjoining pressure $\Pi(r)$. The solution of this more complicated from computational point of view problem will be a goal of future studies. We will conclude this review with a comment that may be instructive for the qualitative analysis of Eq. (19). The main point here is that if the zone where attractive forces act in a heterogeneous wetting film is with dimensions smaller than the critical wavelength from Eq. (14), this film will remain stable with respect to linear perturbations. In the system considered here the attractive zone dimensions are of the order of the bubble contact radius $a$ (Fig. 14), which is equivalent to the fact that linear instability (spontaneous rupture) could be expected if $a \geq \lambda_{cr}$. Nevertheless that for glass/water/air films $\lambda_{cr}$ is of the order of 10 μm (see Fig. 9), i.e. 1000 times larger than a nanobubble, one could hardly expect substantial impact (shortening) on the rupture thicknesses due to instability effects.